\documentclass{kapproc}

\usepackage{procps}

\usepackage{amsmath}
\usepackage{amssymb}

\usepackage[dvips]{graphicx}

\upperandlowercase

\let\footnote\savefootnote

\kluwerbib 

\begin{document}



\articletitle[Rotation, Curvature, and Subcritical Shear
Turbulence]{Subcritical Turbulent Transition in\\
Rotating and Curved Shear Flows}








--------------

\author{Pierre-Yves Longaretti\altaffilmark{1}, Olivier Dauchot\altaffilmark{2}}

\altaffiltext{1}{LAOG, BP 53X, F-38041, Grenoble, France}
\altaffiltext{2}{GIT/SPEC/DRECAM/DSM CEA Saclay, F-91191,
Gif-sur-Yvette, France}


\begin{abstract}
  The effects of global flow rotation and curvature on the
  subcritical transition to turbulence in shear flows are
  examined. The relevant time-scales of the problem are
  identified by a decomposition of the flow into a laminar
  and a deviation from laminar parts, which is performed for
  rotating plane Couette and Taylor-Couette flows. The
  usefulness and relevance of this procedure are discussed
  at the same time. By comparing the self-sustaining process
  time-scale to the time-scales previously identified, an
  interpretation is brought to light for the behavior of
  the transition Reynolds number with the rotation number
  and relative gap width in the whole neighborhood (in
  parameter space) of the non-rotating plane Couette flow
  covered by the available data.
\end{abstract}



\section{Introduction:}

In the last decade or so, a number of breakthroughs have been
achieved in the understanding of the onset of turbulence in
subcritical shear flows, such as the plane Couette flow and
Poiseuille flow, both from an experimental point of view (e.g.,
\cite{Bot98}), and a numerical and theoretical one (e.g.,
\cite{N90}; \cite{CB97}; \cite{HKW95}; \cite{W97}; \cite{W98};
\cite{W03}; \cite{FE03}). In this context, the present
contribution has two main objectives: characterize the effects of
global flow rotation and curvature in subcritical flows from the
available data, and show that these characteristics can be
understood at a semi-quantitative level from time-scale
considerations. Understanding these questions is essential for
geophysical and astrophysical applications, which is one of the
motivations of this work. Data on rotating plane Couette flows and
Taylor-Couette flows are used in this investigation.

Section \ref{adv} is devoted to the identification of rotation and
curvature characteristic quantities, and relating them to the
gross dynamics of the flow. Not surprisingly, the associated
dimensionless numbers reduce to the shear-based Reynolds number,
the rotation number, and the relative gap width (for
Taylor-Couette flows); the novel point is that the relative gap
width is interpreted in terms of a ratio of dynamically relevant
time-scales. The experimental data are then reviewed in section
\ref{data}, and the characterization of the available data in
terms of the previously defined dimensionless numbers is performed
in section \ref{phen}. The last section briefly summarizes the
most important results, and discusses their astrophysical
implications.

\section{The physics of the advection term revisited:}\label{adv}

The main objective of this section is to pinpoint the relevant
time-scales in globally subcritical, rotating and curved flows,
and to relate them to the various contributions of the
advection/acceleration term. This turns out to be essential to
develop a semi-quantitative understanding of the available data on
such flows. In practice, we consider only rotating plane Couette
and Taylor-Couette flows. Incompressibility is assumed throughout.

\subsection{Equations of motions}

The relevant time-scales are well-known in rotating plane Couette
flows, and follow immediately from the expression of the
Navier-Stokes equation in the rotating frame, which reads

\begin{equation}\label{RPC}
  \frac{\partial{\mathbf u}}{\partial t}+{\mathbf u}.{\mathbf\nabla}{\mathbf u}
  =-\frac{{\mathbf\nabla}{\pi}}{\rho}-2{\mathbf \Omega}\times{\mathbf
  u}+\nu\Delta{\mathbf u},
\end{equation}

\noindent with obvious notations (in particular, the centrifugal
term has been included in the pressure gradient). They are the
shear\footnote{The convention adopted here is that the sign of $S$
is chosen to be positive when the flow is cyclonic, i.e., when the
contributions of shear and rotation to the flow vorticity have the
same sign. With the usual choice of axes in plane Couette flows,
this implies that $S=-2S_{xy}$, where $S_{ij}$ is the usual
deformation tensor.} time-scale $t_s=|S^{-1}|$, the viscous one
$t_\nu=d^2/\nu$ ($d$ is the gap in the experiment), and the
rotation time-scale related to the Coriolis force
$t_\Omega=(2\Omega)^{-1}$ ($\Omega$ is the rotation velocity of
the flow in an inertial frame), and relate to the advection term,
the viscous term, and the Coriolis force, respectively.
Correlatively, the flow is described by two dimensionless numbers,
the Reynolds number\footnote{This definition differs from the
usual one by a factor of 4; this convention is adopted here for
consistency with the treatment of Taylor-Couette flows.}
$Re=t_\nu/t_s=|S|d^2/\nu$, and the rotation number $R_\Omega=
\mathrm{sgn}(S) t_s/t_\Omega=2\Omega/S$.

The situation is less straightforward for Taylor-Couette flows,
where the dimensionless number usually associated to the flow
geometry, $\eta=r_i/r_o$ ($r_i$ is the inner cylinder radius,
$r_o$ the outer one) does not obviously correspond to a ratio of
time-scales of the flow. However, on closer inspection, it appears
that this situation arises from a partially incorrect assimilation
of the shear time-scale to the advection term. Indeed, in the case
of solid body rotation, the shear vanishes, while the advection
term does not, due to the global curvature of the flow. One must
therefore devise a way to isolate the global shear contribution to
the advection term from other contributions.

It turns out that one convenient way to operate such a distinction
is to decompose the total flow into its laminar part, and a (not
necessarily small) deviation:

\begin{equation}\label{split}
   {\mathbf u}={\mathbf u}_L+{\mathbf w}.
\end{equation}

\noindent Although the dynamical relevance of the laminar flow to
the turbulent one is not a priori obvious, this procedure is
suggested and justified by the following considerations:

\begin{itemize}
  \item Inasmuch as this is feasible, a distinction between global shear,
  rotation and curvature cannot be operated by a tensorial decomposition
  of the advection term. For example, it is well-known that both a pure
  global shear and a global rotation, such as the ones present in rotating plane
  Couette flows, contribute to the vorticity tensor. In fact, a direct
  Taylor expansion of the deformation for small displacements shows
  that one needs to go at least to second order to distinguish the two
  contributions. Therefore, no tensor constructed from the flow
  velocity first derivatives will establish the required distinction, by
  construction.
  \item The global characteristics of the flow are the same for
  the laminar and turbulent solutions (geometry, global time-scales,
  nature of the boundary condition, etc). Therefore, one way to make
  them appear explicitly in the Navier-Stokes equation is
  precisely to make the proposed decomposition, as the laminar solution
  depends everywhere explicitly on these global characteristics.
  In particular, the laminar and turbulent flows share the same
  boundary conditions (velocity difference on the boundary, gap
  width, etc), so that the relative difference between the laminar
  and turbulent solution is of order unity. This means that the laminar
  solution is a convenient measure of the turbulent one, although their
  detailed mechanisms and characteristics are of course essentially
  different. For example, it turns out that the transition
  Reynolds number is highly sensitive to various global and/or qualitative
  characteristics of the laminar flow, such as time-scales, or
  ``distance''
  in parameter space to the linear stability limits (see section \ref{data}).
\end{itemize}

It is useful to point out where this decomposition leads to for
the rotating plane Couette flow. This is most naturally done in
the rotating frame, so that Eq.~(\ref{RPC}) becomes


\begin{equation}\label{split-RPC}
  \frac{\partial{\mathbf w}}{\partial t}+
  {\mathbf w}.{\mathbf\nabla}{\mathbf w}=
  S\cdot y \frac{\partial{\mathbf w}}{\partial x}
  + (2\Omega+S) w_y {\mathbf e}_x
  - 2\Omega w_x {\mathbf e}_y
  -\frac{{\mathbf \nabla}\delta\pi}{\rho}+\nu\Delta{\mathbf w},
\end{equation}

\noindent where the pressure gradient balancing the laminar flow
Coriolis force has been subtracted out to form the effective
generalized pressure $\delta\pi$. Note that on the walls, which
specify the global characteristics of the flow, the boundary
condition becomes ${\mathbf w}=0$, quite a featureless constraint:
the effect of these boundaries is now explicitly included in the
Navier-Stokes equation through the dynamical linear forcing terms
on the right-hand side. The real usefulness of this change of
point of view comes out when considering Taylor-Couette flows, as
we shall now argue.

In this flow, the laminar solution takes the form ${\mathbf
u}_L({\mathbf r})=r\Omega(r){\mathbf e}_\theta$. Note that the
rotating plane Couette flow can be viewed as a limit of small
relative gap width $d/\bar{r}\rightarrow 0$ ($\bar{r}$ is some
characteristic radius of the flow), at constant shear, and
constant rotation. Then, the effect of the global curvature and
rotation will be more easily distinguished from one another if one
chooses a formulation of the Navier-Stokes equation which makes
the difference with Eq.~(\ref{split-RPC}) explicit. To this
effect, one must define a characteristic rotation velocity
$\bar{\Omega}$, and a characteristic shear\footnote{The shear of
the laminar flow is defined as $S=rd\Omega/dr=2S_{r\phi}$ in order
to maintain the sign convention adopted for rotating plane Couette
flows for cyclonic and anticyclonic rotation.} $\bar{S}$ of the
laminar flow. A convenient way to do this is to choose a
characteristic radius $\bar{r}$, and impose
$\bar{\Omega}=\Omega(\bar{r})$, $\bar{S}=S(\bar{r})$; the choice
of $\bar{r}$ does not need to be further specified for the time
being (this point is discussed in the next subsection). Defining
$\delta\Omega=\Omega(r)-\bar{\Omega}$ and $\delta S=S(r)-\bar{S}$,
the decomposition of the Navier-Stokes equation of the
Taylor-Couette flow leads to ($r,\phi,z$ is the coordinate system
in the rotating frame)

\begin{align}\label{split-TC}
  \frac{\partial{\mathbf w}}{\partial t}+
  {\mathbf w}.{\mathbf\nabla}{\mathbf w}= &
  -\delta\Omega \partial'_\phi{\mathbf w}
  - (2\bar{\Omega}+\bar{S}) w_r {\mathbf e}_\phi
  + 2\bar{\Omega} w_\phi {\mathbf e}_r
  -\frac{{\mathbf \nabla}\delta\pi}{\rho}+\nu\Delta{\mathbf w}\nonumber \\
  & - (2\delta\Omega+\delta S) w_r {\mathbf e}_\phi
  + 2\delta\Omega w_\phi {\mathbf e}_r,
\end{align}

\noindent where one has used $\partial_\phi{\mathbf
w}=\partial'_\phi{\mathbf w} + \delta\Omega w_r {\mathbf e}_\phi
  + 2\delta\Omega w_\phi {\mathbf e}_r$, with
$\partial'_\phi{\mathbf w} \equiv(\partial_\phi w_r){\mathbf e}_r
+(\partial_\phi w_\phi){\mathbf e}_\phi + (\partial_\phi
w_z){\mathbf e}_z$; this definition is introduced so that the
contributions of order $1/\bar{r}$ of the derivatives in the
linear terms are separated from the contributions of order $1/d$.
This equation is similar\footnote{In the identification of the two
equations, note that $r\longleftrightarrow y$ and
$\phi\longleftrightarrow -x$. Also, $\delta\Omega\simeq
\bar{S}(r-\bar{r})/\bar{r}$, an approximation which holds to $\sim
10\%$ for the range of $\eta$ explored in the available
experiments, a feature needed in the comparison of
Eqs.~(\ref{split-RPC}) and (\ref{split-TC}).} to
Eq.~(\ref{split-RPC}), except for the last two terms, which
consequently are connected to the global flow curvature. Note
that, although the definition of $\delta\Omega$ and $\delta S$
depends on the choice of $\bar{r}$, the overall variation of these
quantities throughout the flow, $\Delta (\delta\Omega)$ and
$\Delta(\delta S)$, does not. In fact, $|\Delta(\delta\Omega)|
\sim |\Delta(\delta S)| \sim |\bar{S}|d/\bar{r}$, as can be
checked from the laminar flow profile $\Omega(r)=A+B/r^2$. In the
process, four time-scales of the incompressible Taylor-Couette
flow have been identified: they are the shear time-scale
$t_s=|\bar{S}^{-1}|$, the rotation time-scale
$t_\Omega=(2\bar{\Omega})^{-1}$, a curvature-related time-scale
that one can define as $t_{\mathcal C}=|\bar{S}^{-1}|\bar{r}/d$,
and, of course, the viscous time-scale $t_\nu=d^2/\nu$. This also
shows that the dimensionless geometric number $\eta$ can be
related to the ratio of the shear and curvature time-scales.
Taking the limit $d/\bar{r}\rightarrow 0$ ($\eta\rightarrow 1$),
one recovers the rotating plane Couette relation
Eq.~(\ref{split-RPC}).

\subsection{Characteristic quantities, dimensionless numbers, and the
curvature and rotation concepts:}

Taylor-Couette flows possess three dimensionless control
parameters, which are usually chosen as the Reynolds numbers
associated to the inner and outer cylinder rotation velocities,
$R_i$ and $R_o$, and $\eta$, the ratio of their two radii. The
preceding discussion suggests that one should use instead ratios
of time-scales, which have a more direct dynamical meaning. This
defines the shear-based Reynolds number\footnote{We use the same
Reynolds number definition for rotating plane Couette and
Taylor-Couette flows, based on the total gap width and total
velocity difference. Consequently, the quoted Reynolds numbers for
plane Couette flows differ from the ones in the literature by a
factor of 4.} $Re=t_\nu/t_s$, the rotation number
$R_\Omega=\mathrm{sgn}(S) t_s/t_\Omega$, and a ``curvature" number
$R_{\mathcal C}=t_s/t_{\mathcal C}$. Once a choice of $\bar{r}$ is
operated (see below), the procedure is completely specified. This
three-dimensional parameter space, in which notable curves and
surfaces are drawn, is represented in figure \ref{pylplot}.

However, the physical meaning of this procedure is less
straightforward than one would like, and this is related to an
obvious weakness of Eq.~(\ref{split-TC}): the distinction between
the rotation and curvature terms is not absolute when the related
time-scales are both dynamically significant. Indeed, any change
of definition of $\bar{\Omega}$ and $\bar{S}$ results in a
correlative change of $\delta\Omega$ and $\delta S$. Nevertheless,
the physical meaning is to a large extent unambiguous in at least
two different contexts:

\begin{itemize}
  \item If one changes the rotation velocity of the inner and
  outer cylinders by the same quantity, this will change
  $\bar{\Omega}$ by the same amount (independently of the choice of $\bar{r}$),
  while leaving all other quantities ($\bar{S}$, $\delta\Omega$, $\delta S$) unchanged.
  Such changes are obviously an effect of changes in the flow
  rotation.
  \item On the other hand, when changes in the flow are operated while
  maintaining $t_\Omega\lesssim t_{\mathcal C}$, the physical
  meaning of the distinction between the characteristic quantities ($\bar{\Omega}$,
  $\bar{S}$) and the deviations from these ($\delta\Omega$, $\delta S$)
  is blurred. This is the case in particular when the cylinders
  are counter-rotating, or when one cylinder is at rest, for \textit{any}
  choice of $\bar{r}$. In such a context, changes in both parameters (the
  rotation number and the curvature number) describe the effect of a
  change in the flow curvature, as they are both proportional to $d/\bar{r}$,
  and vanish if the limit of vanishing global curvature is taken while
  enforcing the $t_\Omega\lesssim t_{\mathcal C}$ relation.
\end{itemize}

This clearly shows that rotation and curvature are not
interchangeable concepts, although they have a non-negligible
overlap. In this context, the denominations ``rotation number" and
``curvature number" are somewhat conventional and partially
misleading, even if justified to some extent by the preceding
considerations. In the ($R_\Omega$, $R_{\mathcal C}$) plane,
changes along lines of constant $R_{\mathcal C}$ correspond to
changes of rotation, but there are infinitely many paths involving
changes of both parameters and corresponding to changes of
curvature from a physical point of view. Furthermore, most paths
do not lead to any clear-cut distinction between curvature and
rotation changes in the flow. Of course, once a reference
curvature path is chosen (e.g., the path with the inner cylinder
at rest, corresponding to the inviscid linear stability limit on
the cyclonic side, shown as curve (4) on figure \ref{pylplot}),
every point in the ($R_\Omega$, $R_{\mathcal C}$) plane can be
connected to the non-rotating plane Couette flow (the origin in
the plane), first through a change of curvature along the chosen
curvature path until the desired curvature number is reached, and
then through a change of rotation at constant $R_{\mathcal C}$.
However, this distinction is only relative.

Obviously, this situation is intrinsic, as one cannot curve a
straight flow, without at the same time making it rotate. The
procedure outlined here nevertheless leads to the definition of
well-defined parameters, which have a dynamical interpretation. It
is their physical meaning in terms of rotation and curvature which
is partially ambiguous. Furthermore, these parameters turn out to
be useful to understand basic features in the data on the
subcritical transition to turbulence, as discussed in the next
section.

The remaining point to be addressed relates to the choice of
$\bar{r}$. The preceding discussion makes it clear that this
choice is not unique. The definition we have adopted here is
$\bar{r}\equiv (r_i r_o)^{1/2}$, as suggested in \cite{DB04}
($r_i$ and $r_o$ are the inner and outer radius, respectively).
This choice is partially motivated by the compactness of the
resulting expressions for the dimensionless numbers introduced
above:

\begin{equation}\label{Re}
Re= \frac{2\bar{r}}{r_o+r_i}\frac{\bar{r}|\Omega_o-\Omega_i|
d}{\nu} =\frac{2}{1+\eta}|\eta R_o - R_i|,
\end{equation}

\begin{equation}\label{Rom}
R_\Omega=\frac{r_i\Omega_i+r_o\Omega_o}{r_o
r_i}\frac{d}{\Omega_o-\Omega_i} =(1-\eta )\frac{R_i+R_o}{\eta R_o
- R_i},
\end{equation}

\begin{equation}\label{Rc-eta}
R_{\mathcal C}=\frac{d}{\bar{r}}=\frac{1-\eta}{\eta^{1/2}}.
\end{equation}

Note that, for the range of values of $\eta$ explored in the
available experiment, $Re\simeq \bar{r}|\Delta\Omega| d/\nu$, and
$R_\Omega \simeq (2\bar{\Omega}/\bar{r})(d/\Delta\Omega)$ (within
a few percents).

\section{Subcritical transition in rotating plane Couette and
Taylor-Couette flows:}\label{data}

Considering a laminar flow of given dimensionless numbers
(rotation, curvature, and Reynolds), two different things can
happen when increasing the Reynolds number: either the flow will
undergo a linear instability first (supercritical transition), or
it will undergo a laminar-turbulent transition first (globally
subcritical transition). The second option may happen whether the
flow is linearly unstable or not.

The Reynolds number characterizing subcritical transition in a
system is not a quantity that can be measured with absolute
precision, as it depends to some extent on the experimental
protocol used in its determination. For example, the
laminar-turbulent transition Reynolds number is generically larger
than the turbulent-laminar one. Furthermore, the flow is
intermittent over a range of Reynolds numbers in the vicinity of
this transition. This leads to some differences in the determined
Reynolds transition values, even when the same data are used by
different authors; however, the dispersion of the data in a given
author's choice is much smaller. Overall, the resulting range of
values (at given dimensionless numbers) is uncertain within less
than a factor of $\sim 2$; we shall ignore this problem here, as
we are only interested in characterizing qualitative trends and
orders of magnitude.

With this convention, both supercritical ($Re=R_c$) and globally
subcritical ($Re=R_g$) transitions are characterized by surfaces
in the three dimensional space ($R_\Omega, R_{\mathcal C}, Re$).
Only particular lines on these surfaces have been probed by the
available experiments. Obviously, the supercritical and
subcritical surfaces meet somewhere in this space, so that one
needs to characterize both surfaces.

\begin{figure}[htb]
\centering
\includegraphics[scale=0.7]{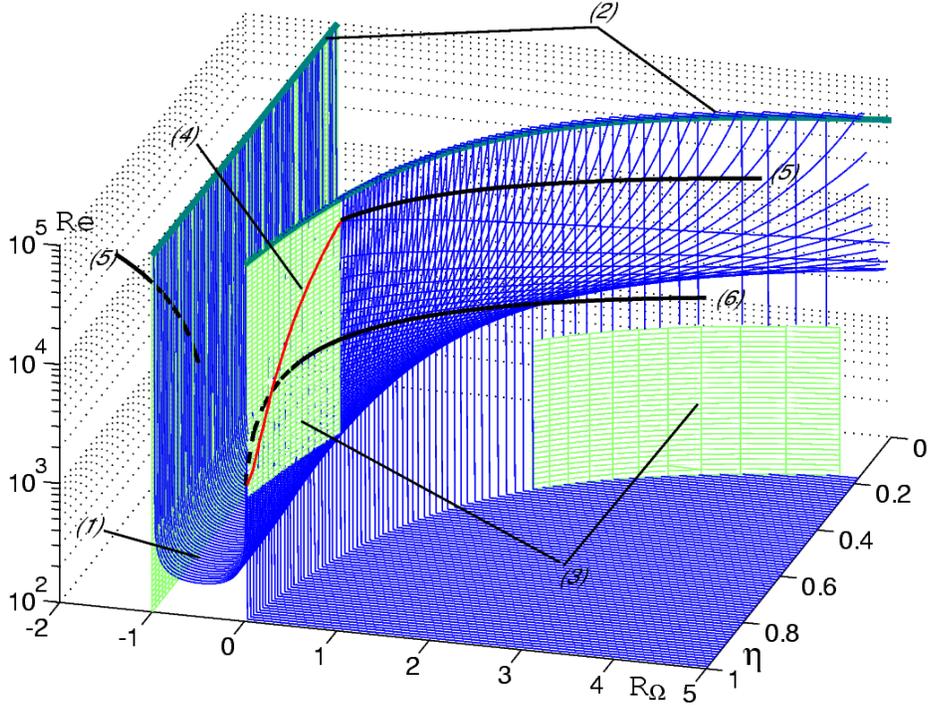}
\caption{\small Stability boundaries in the $(Re, R_\Omega,
R_{\mathcal C})$ parameter space ($\eta$ is used as a measure of
$R_{\mathcal C}$ on this graph).Manifold (1): linear stability
threshold. Dark curves (2): linear stability threshold in the
inviscid limit. Manifold (3): extrapolation of the inviscid
criteria throughout the full Reynolds space (partially shown for
readability). Curve (4): globally subcritical threshold obtained
with inner cylinder at rest (see text). Other curves: globally
subcritical thresholds obtained at fixed value of $\eta$ [(5):
$\eta=0.7$; (6): $\eta=1$] (see text).}\label{pylplot}
\end{figure}

It turns out, from a practical point of view, that the
supercritical surface (manifold (1) in figure \ref{pylplot}) is
sufficiently well captured by the analytic approximation derived
by \cite{EG96}, for Taylor-Couette flows, although it relates only
to axisymmetric perturbations (in other words, the
non-axisymmetric perturbations seem to play little role in the
definition of the supercritical transition). The rotating plane
Couette flow is included in the limit $\eta\rightarrow 1$. For the
relatively high Reynolds numbers of interest for subcritical
transitions to turbulence ($> 1000$), the dependence of the
supercritical surface on the Reynolds number is very steep, and
the surface is well-approximated by the inviscid linear stability
limit (curves (2) of figure \ref{pylplot}). This explains that the
limit of the subcritical regime in the ($R_\Omega, R_{\mathcal
C}$) plane is well-approximated by the inviscid limit. Linear
instability follows somewhere in the fluid in this limit if

\begin{equation}\label{linstab}
  -1 \leqslant \frac{2\Omega(r)}{S(r)} \leqslant 0,
\end{equation}

\noindent at this location (this is equivalent to the constraint
put by the Rayleigh discriminant). Asking that the fluid is
everywhere stable with respect to this criterion translates into
$R_\Omega < -1$, or $R_\Omega > (1-\eta)/\eta$.

\begin{figure}[htb]
\centering
\includegraphics[scale=1.15]{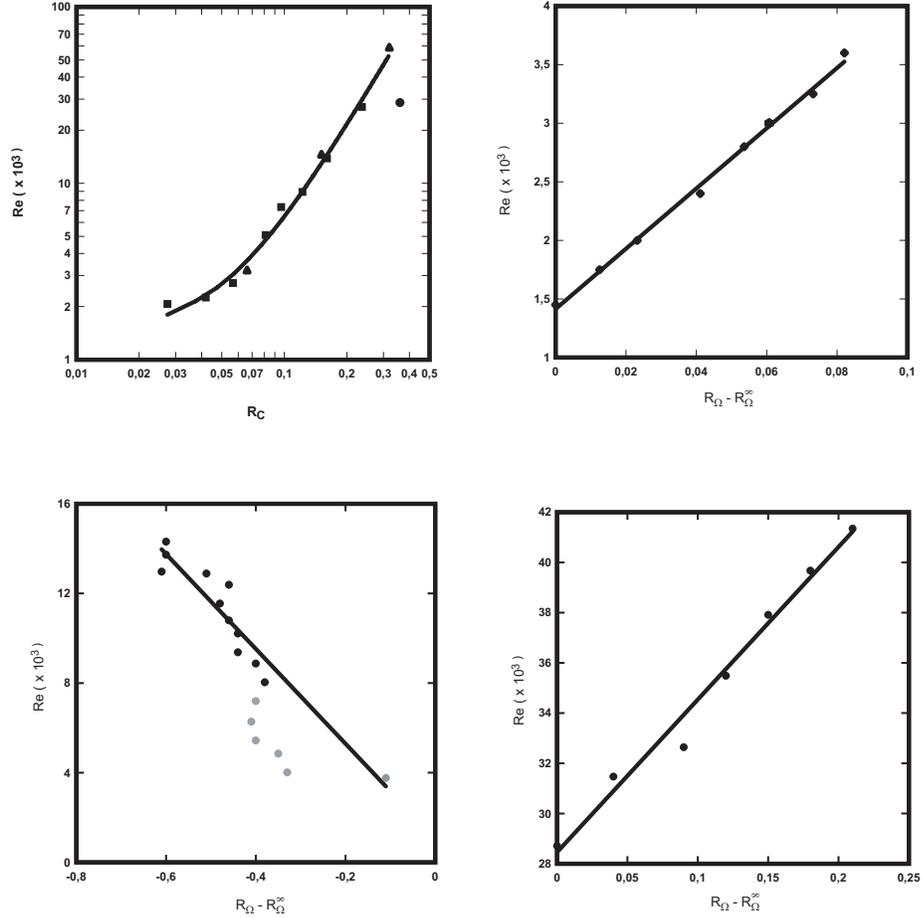}
\caption{\small Data on subcritical transition. Upper left: data
of Wendt (1933) and Taylor (1936), for cyclonic rotation, with the
inner cylinder at rest, for varying $\eta$; this shows the
dependence on the curvature number in cyclonic flows. Upper right:
data of Tillmark and Alfredsson (1996) for rotating plane Couette
flows ($\eta=1$). Lower left and lower right: data of Richard
(2001), for anticyclonic and cyclonic rotation, respectively
($\eta=0.7$). On the last three graphs, the curvature number is
held fixed, so that the data show the dependence of the transition
Reynolds number on the rotation number. The solid lines represent
best quadratic or linear fits to the data.
$R_\Omega^\infty=(1-\eta)/\eta$ for cyclonic flows, and
$R_\Omega^\infty=-1$ for anticyclonic ones.}\label{dataplot}
\end{figure}

The data on subcritical transition discussed here are those of
Wendt (1933), Taylor (1936), Tillmark and Alfredsson (1996), and
Richard (2001); they are represented on figure \ref{dataplot}. One
could also include data on counter-rotating cylinders
(\cite{And86}; \cite{Prig03}), but these occupy only a small area
in the ($R_\Omega, R_{\mathcal C}$) plane, and bring little
constraint on the trend of the transition Reynolds number with the
rotation and curvature numbers (see \cite{LD04} for a discussion o
couter-rotating data).

The data are well parameterized by the following approximate
formula (the $+$ and $-$ sign refer to cyclonic and anticyclonic
flows, respectively)

\begin{equation}\label{Rgpm}
  R_g^{\pm}(R_\Omega, R_{\mathcal C})=R_{PC}^{\pm} + a^{\pm}(\eta)|R_\Omega -
  R_\Omega^{\infty,\pm}| + b^{\pm} R_{\mathcal C}^2,
\end{equation}

\noindent with $R_{PC}^{+}\simeq 1400$, $R_{PC}^{-}\simeq 1100\sim
R_{PC}^+ $, $21000 \lesssim a^\pm \lesssim 61000$, $2\times 10^5
\lesssim b^+\lesssim 6\times 10^5$, $b^- \ll b^+$, and where
$R_\Omega^{\infty,+}=(1-\eta)/\eta$, and $R_\Omega^{\infty,-}=-1$.
This is discussed in detail in \cite{LD04}, and \cite{DB04}.

The most notable characteristics of this dependence are the
following:

\begin{itemize}
  \item The linear dependence on the rotation number, and
  quadratic one on the curvature number.
  \item The rather steep dependence with both numbers ($a^{\pm}$ and $b^+$
  are large numbers).
  \item The apparent symmetry between cyclonic and anticyclonic
  rotation number dependence, and the absence of dependence on the
  curvature on the anticyclonic side ($b^-\simeq 0$).
\end{itemize}

The next section discusses these features; in particular, the
strength of this dependence is explained in order of magnitude on
the basis of time-scale considerations.

\section{Data interpretation:}\label{phen}

The linear and quadratic dependence just pointed out can be viewed
as the result of a Taylor expansion. For cyclonic rotation, this
expansion is performed around the non-rotating plane Couette flow,
first in terms of $R_{\mathcal C}$ along curve (4) on figure
\ref{pylplot}, and then away from it, at constant $R_{\mathcal
C}$, in terms of $R_\Omega$. For anticyclonic rotation, the
expansion depends only on $R_\Omega$, at least for $\eta > 0.7$,
and is performed from the marginal stability state, at
$R_\Omega\simeq -1$.

\subsection{Linear dependence}

The linear dependence with $R_\Omega$ is neater on the cyclonic
data than on the anticyclonic one, but the data extend less far on
the cyclonic side, unfortunately. The increased dispersion on the
anticyclonic side has several reasons:

\begin{itemize}
  \item Both cylinders need to be rotating at much higher speed than on
  the cyclonic side to reach the subcritical turbulent transition. This
  automatically reduces the precision of the measurements.
  \item The quantity $R_\Omega$ amplifies the uncertainty due to small errors
  in the determination of the cylinder angular velocities at
  transition.
  \item There is an important intrinsic difference between the
  cyclonic and anticyclonic marginal stability limits. On the
  cyclonic side, instability begins at a single location (the
  inner radius), whereas on the anticyclonic side, marginal
  stability applies throughout the fluid (this follows because the
  constant angular momentum solution is a laminar solution of the
  Taylor-Couette flow). Equivalently $2\delta\Omega(r)+\delta
  S(r)=0$ at the anticyclonic marginal stability limit. Therefore,
  the fluid is much more sensitive to a potential linear
  instability. In particular, the unavoidable Eckmann circulation
  can much more easily make the criterion Eq.~(\ref{linstab})
  satisfied somewhere in the flow on the anticyclonic side than on the
  cyclonic one. We believe that this feature most likely explains why
  Richard's data show only a weak dependence of the transition Reynolds
  number on the rotation number out to $|1+R_\Omega|\simeq 0.35$, and then a
  sharp increase to reach back the linearly varying regime. The
  related data points are shown as thin dots in the lower left
  quadrant of Fig.~\ref{dataplot}, and not used in the linear fit.
\end{itemize}

Note that the mutual cancellation of a part of the curvature terms
in Eq.~(\ref{split-TC}) at the anticyclonic marginal linear
stability limit just pointed out also provides a natural
explanation for the apparent absence of dependence of the
anticyclonic data on the curvature number.

\subsection{Quadratic dependence}

The quadratic dependence of the data on the gap width when keeping
the inner cylinder at rest has already been pointed out by
\cite{Zel81}, and \cite{RZ99}. At least three explanations of this
behavior have been put forward in the literature (\cite{Zel81};
\cite{Bulle93}; \cite{long02}). These will be commented below.

Let us first note that, in plane Couette flows, transition occurs
at a constant Reynolds number $Re=\bar{S}d^2/\nu$. Therefore, at a
given shear $\bar{S}$, the scale $d$ is the characteristic scale
of turbulence in such a system. Conversely, in the quadratic
asymptotic regime, [$Re\propto (d/\bar{r})^2$], transition occurs
at constant $Re^*=\bar{S}\bar{r}^2/\nu$ ($=b^+$), a point already
made by \cite{RZ99}. Consequently, the characteristic radius
$\bar{r}$ instead of the gap $d$ characterizes the transition at a
given shear. This unambiguously points out curvature and not
rotation as the source of this behavior, consistently with the
discussion of section 2.

\cite{Bulle93}, explains the quadratic behavior by considering the
growth of finite amplitude local defects in the laminar profile.
However, only WKB modes of instability created by the defects are
considered, for which the scale $\bar{r}$ cannot play any role.
This is why we believe that this analysis cannot capture the
transition mechanism.

\cite{Zel81}, phenomenological explanation is based on the
following two ideas:

\begin{itemize}
  \item The transition Reynolds number may depend on the single
  time-scale ratio $Ty(r)=4\kappa^2(r)/S^2(r)$ at some appropriately
  chosen radius in the flow: $R_g=f(Ty)$; $\kappa(r)=[2\Omega(2\Omega+S)]^{1/2}$
  is the epicyclic frequency, i.e., the frequency of oscillation
  of the whole fluid, under the combined action of the shear and
  the Coriolis force.
  \item A ``split-r\'egime" of instability may occur, in
  which the inner portion of the flow undergoes a transition to
  turbulence at lower Reynolds number than the whole flow, once
  a large enough relative gap width is reached.
\end{itemize}

Considering the relative ambiguity in the distinction between the
rotation and curvature time-scale discussed in section \ref{adv},
one may indeed ask whether a single time-scale would be sufficient
to understand the data. This would imply that Taylor-Couette flows
possess a hidden redundancy, and could be described by two
appropriately chosen parameters, and not three. However, the
extended set of data used here does not support this idea,
although a larger body of experimental results is probably needed
to ascertain this result.


\cite{long02} has developed an alternate phenomenology of
subcritical shear flow turbulence. It relies on the one hand on a
turbulent viscosity description, in which the characteristic
length is identified to the top of the Kolmogorov cascade; on the
other hand it makes use of the constraint that a subcritical shear
flow is out of thermodynamic equilibrium and tries to restore
equilibrium by transporting momentum across the shear, and, to do
so, chooses the most efficient of the two means at its disposal
(laminar or turbulent transport). This provides scaling laws for
the characteristic length and velocity of turbulent eddies, by
relating them to the Reynolds number of transition to turbulence.
By further noting that for large enough gap widths, the turbulent
eddies must unavoidably scale with the radius and not the gap, the
quadratic r\'egime is recovered.


\subsection{Orders of magnitude}

The last point we wish to address is the origin of the large
values the coefficients $a$ and $b$ which appear in the
characterization of the data performed in Eq.~(\ref{Rgpm}). We
mostly consider cyclonic flows, for which the physics is best
understood

As mentioned in the introduction, an important breakthrough in the
understanding of subcritical turbulence in non-rotating plane
Couette flow comes from the work of \cite{W97}, who analyzed by
the means of quasi-linear theory a turbulent self-regeneration
process previously observed in the numerical simulations of
\cite{HKW95}. These last authors have tracked down turbulence to
the smallest unit where it is self-sustained by reducing
appropriately the simulation box size and the Reynolds number. A
very important feature of the identified self-sustaining process
is that it has a rather long time-scale compared to the shear:

\begin{equation}\label{SSP}
t_{ssp}^+\sim 100 \bar{S}^{-1}.
\end{equation}
\noindent

\noindent and that the scales involved in the self-regeneration
mechanism are comparable to the flow width. This time-scale is the
shortest of all the mechanisms found in the systematic reduction
of the flow, and thus corresponds to the most robust one, which
involves two streamwise rolls in the spanwise direction. These
streamwise rolls, first observed by \cite{DD95}, typically scale
on the gap width. Accordingly, it is very likely that such a long
time-scale is a generic feature in non rotating plane Couette
flows, because of the large ($R_{PC}^+\sim 1500$) Reynolds number
of transition to turbulence which are always observed in these
systems. Such a large Reynolds number constrains the viscous
diffusion time at the scales involved in the self-regeneration
mechanism. Typically for a length-scale $d/4$, the equality of the
viscous time $t_\nu=(d/4)^2/\nu$ with  $t_{ssp}^+$ would indeed
leads to $R_g=1600$. Such a scale ($d/4$) is characteristic of the
thickness of the streaks, apparently the smallest characteristic
scale of the process.

From a physical point of view, one expects that the contributions
of either rotation or curvature become comparable to $R_{PC}^+$ in
Eq.~(\ref{Rgpm}) when the rotation or curvature time-scales
decrease to become comparable to $t_{ssp}^+$. Correlatively, the
fact that this happens when the rotation and curvature numbers are
$\simeq 1/20$ comes as no coincidence. Indeed, the timescale
associated with the Coriolis term is $t_\Omega\simeq
(2\bar{\Omega})^{-1}$, while the timescale associated with the
curvature terms is $t_{\mathcal C}\simeq (\delta\Omega)^{-1}\sim
\bar{S}\bar{r}/d$. They become comparable to $t_{ssp}^+$ as
defined in Eq.~(\ref{SSP}) when $R_\Omega$ or $R_{\mathcal C}$
exceeds $10^{-2}$, which is remarkably close to the actual value
of $1/20$, considering the qualitative nature of the argument.
This physical constraint is also what primarily determines the
magnitude of $a^+$ and $b^+$ (once the form of the dependence on
$R_\Omega$ and $R_c$ is known). Indeed, in rotating plane Couette
flows, requiring $a^+ R_\Omega \gtrsim Re_{PC}^+$ when $\ R_\Omega
\gtrsim 1/20$, implies that $a^+ \sim 10^4$. Similarly, in
Taylor-Couette flows, requiring that $b^+ R_{\mathcal C}^2 \gtrsim
Re_{PC}^+$ when $\ R_{\mathcal C} \gtrsim 1/20$ leads to $b^+ \sim
10^5$.

One can see that the rather large values of the transition
Reynolds number, as well as of the coefficients characterizing the
effect of rotation ($a^+$) and curvature ($b^+$) can be ascribed
to a single origin: the rather large ratio of the turbulence
self-regeneration process to the shear time scale. Still, we do
{\it not} infer that the self-sustaining mechanism proposed by
Waleffe is valid in the presence of rotation. We just use the fact
that in order to modify or even suppress this mechanism, the
rotation or curvature effects must have timescales of the same
order.  On the contrary, the above analysis clearly indicates that
a better understanding of this process in the framework of curved
shear flows, and identifying it in the presence of rotation, is of
primary importance for future progress.

The self-sustaining process has not yet been identified at the
anticyclonic marginal stability limit, and its nature is not
known. However, one can reverse the reasoning expressed right
above to reach the conclusion that its characteristic time-scale
is also $\sim 100 \bar{S}^{-1}$. Testing this conjecture would
bring support to the framework developed here to analyze and
understand the data.

\section{Conclusions and implications:}\label{conc}

We feel that this work brings to light a few important points,
which we believe to be of potentially more general applicability
than what was done here:

\begin{itemize}
  \item It is both meaningful and useful to decompose the
  Navier-Stokes equation into a laminar part of the flow, and a
  deviation from the laminar flow. This helps identifying the
  relevant time-scales in the flow, and isolating which ``portion"
  of the advection term is directly related to the physics
  described by the Reynolds number. The same procedure is
  also useful to relate various flows to one another.
  \item The general trends and features of the transition Reynolds
  number data can be understood with the help of the previously
  identified time-scales, and over a reasonably extended fraction
  of the parameter space, once the turbulence self-sustaining
  process time-scale is identified at one point in the parameter
  space. The procedure in turns constrains to some extent the
  time-scale of the self-sustaining process in the domain where
  data are available, a point we have only briefly touched upon in the
  preceding section.
\end{itemize}

In the process, we have tried to elucidate a little more the
relation between the flow global rotation and its global
curvature. We found that rotation and curvature effects most
probably cannot be accounted for by a single time-scale, as
suggested by \cite{Zel81}, but that two time-scales are required,
in accordance with the fact that Taylor-Couette flows are
characterized by two dimensionless numbers besides the Reynolds
number.

The data described here, and the analysis we have developed, have
some bearing to a related astrophysical problem, namely, the
existence of subcritical turbulence in keplerian accretion disks,
a question which has been the object of an important debate in the
astrophysical community over the three or four past decades. Such
disks are observed in relation to the formation of young stars;
they are also believed to be present in a variety of other
astrophysical objects, such as active galactic nuclei,
microquasars, and so on. A large scale keplerian profile
($\Omega(r)\propto r^{-3/2}$) follows if the disk is cold enough.
The profile can nevertheless stochastically deviate from keplerian
on scales comparable to the disk scale height $h \ll r$. Young
(proto-)star disks are also probably not ionized enough for MHD
processes to be relevant over a significant fraction of their
extent.

The microscopic transport in these disks is known to be many
orders of magnitude smaller than the one inferred from the
observations, so that these disks are believed to be turbulent.
Hydrodynamic Keplerian disks are linearly stable in their most
simple flavors. It was believed until the mid-90's that they were
nevertheless hydrodynamically turbulent, on the basis of the
experimental evidence of subcritical turbulence in non-rotating
Couette flows.

This belief was challenged by the numerical simulations of
\cite{BHS}, and \cite{HBW}. These authors have studied whether
keplerian disks are locally turbulent, by reducing the
Navier-Stokes equation in the disk to Eq.~(\ref{split-RPC}), with
``shearing-sheet'' boundary conditions (they ignore the disk
vertical stratification), thus asking the question in a framework
which is extremely close to the one studied here. They found that
a dynamically significant and stabilizing Coriolis force prevents
the appearance of turbulence for keplerian-like flows
($R_\Omega=-4/3$), up to to the highest resolution achieved in the
simulation ($256^3$ with finite-difference codes; see the
referenced papers for detail).

These simulations have had a very large impact in the astrophysics
community, where the now most largely spread opinion is that a
linear instability is needed for turbulence to show up, a somewhat
excessive position. However, the experimental results of
\cite{Rich01}, seem to imply that keplerian-like flows should be
turbulent at very modest (in astrophysical standards) Reynolds
numbers. The problem is currently reinvestigated from a numerical
point of view (Lesur and Longaretti, \textit{in preparation}).
Preliminary results indicate that the Coriolis force does not
prevent the existence of turbulence in rotating plane Couette
flows, but of course alters the turbulence properties, as one
would expect. However, the relevance of subcritical hydrodynamic
turbulence to accretion disk transport remains to be more
precisely investigated.






\end{document}